\documentclass[pra,aps,10pt,twocolumn,fleqn,floatfix,showpacs,showkeys]{revtex4-1}
\usepackage{graphicx}
\usepackage{amssymb}
\usepackage{amsmath}
\usepackage{bm}
\newcommand\ve[1]{\mbox{\boldmath $#1$}}

\begin{document}
\title{Reentrant topological transitions with Majorana end states in 1D superconductors by lattice modulation}
\author{Masaki \surname{Tezuka}}
\email{tezuka@scphys.kyoto-u.ac.jp}
\author{Norio \surname{Kawakami}}
\affiliation{Department of Physics, Kyoto University, Kyoto 606-8502, Japan}
\date{\today}
\begin{abstract}
The possibility to observe and manipulate Majorana fermions as end states
of one-dimensional topological superconductors has been actively discussed recently.
In a quantum wire with strong spin-orbit coupling placed in proximity to a bulk superconductor,
a topological superconductor has been expected to be realized when the band energy is split
by the application of a magnetic field.
When a periodic lattice modulation is applied multiple topological superconductor phases appear in the phase diagram.
Some of them occur for higher filling factors compared to the case without the modulation.
We study the effects of phase jumps and argue that the topologically
nontrivial state of the whole system is retained even if they are present.
We also study the effect of the spatial modulation in the hopping parameter.
\end{abstract}
\maketitle
\section{Introduction}
Much work has been devoted to realize topologically nontrivial states of matter
with topologically protected surface states.
Majorana surface states are expected to be formed at the boundaries and vortex cores
of topological superconductors (TSs).\cite{Read2000, Kitaev2001, Ivanov2001, Fu2008, Fujimoto2008, Sato2009, Linder2010}
Such states have been expected in a one-dimensional (1D) quantum wire with spin-orbit interaction (SOI)
placed under a Zeeman field and in proximity to a bulk superconductor,\cite{Sau2010, Alicea2010, Lutchyn2010, Oreg2010}
in cold atoms in optical lattices with effective gauge fields generated by spatially varying laser fields,\cite{Sato2009}
bilayer electron gases in semiconductor heterostructure with interlayer Coulomb coupling,\cite{Nakosai2012} among others.
Several groups of experimentalists have recently reported that
they have observed the signatures of Majorana fermions appearing at the ends of nanowires attached to superconductors.\cite{Mourik2012, Rodrigo2012, Deng2012, Rokhinson2012}
For a review we refer to Ref. \onlinecite{Stanescu1302.5433}.

The effect of spatial inhomogeneity on a superconducting quantum wire
has been a nontrivial problem.\cite{Motrunich2001,Gruzberg2005,Brouwer2011,Lutchyn2012,Lobos2012,Bagrets2012,Takei2012,Adagideli2013,Sau2013}
The energy distributions of the end states have been obtained for
the Dirac equation with random mass and a 1D spinless superconductor.\cite{Brouwer2011}
The interplay of disorder and correlation in 1D TSs has also been investigated.\cite{Lobos2012}
Among the possible realizations of spatial inhomogeneity,
quasiperiodic (Harper) potential modulation forms a special class in that
in 1D system all the single particle eigenstates become localized at the same modulation strength.
Quasiperiodic potentials have been experimentally studied in optically trapped cold atom systems \cite{QuasiperiodicColdAtoms}
as well as in solid state, misfit compound \cite{1301.6888} systems.
Signatures of a Hofstadter butterfly-like band structure have been observed in 
a van der Waals system of monolayer graphene on top of a hexagonal boron nitrade surface.\cite{Hunt1303.6942}

The 1D effectively spinless superconductor, expected in 1D superconductor with strong spin-momentum coupling
under magnetic field, is of the D symmetry class.\cite{PeriodicTable}
Therefore its topology is classified by a $Z_2$ topological number.
This suggests that the boundary between two topologically non-trivial 1D superconductors
would not have a localized mode.
Boundary phenomena between topologically equivalent or distinct phases, with Harper or Fibonacci potentials,
have been experimentally studied using photonic quasicrystals.\cite{VerbinPRL2013}

In the scenario of Refs. \onlinecite{Sau2010, Alicea2010, Lutchyn2010, Oreg2010}
the chemical potential needs to lie close to the band edge so that the band degeneracy is removed by the
external magnetic field.
However, the present authors have observed that, by a quasiperiodic lattice modulation with a fixed wavenumber,
effectively single-band superconductor with end Majorana fermions is realized even when the chemical potential is
closer to the center of the original cosine band, because energy separations are introduced within each of the Zeeman-split bands.\cite{Tezuka2012}
We have also demonstrated that this physics is stable even in the presence of a Hubbard-like on-site interaction and/or
a harmonic trapping potential.
More recently, the effect of incommensurate potentials on 1D p-wave superconductors have been studied.\cite{DeGottardi1208.0015, Cai1208.2532, Satija1210.5159}
\textit{Commensurate} diagonal or off-diagonal Harper model has also attracted theoretical attention.\cite{Lang1207.6192, Ganeshan1301.5639}

Here we are interested in characterizing the new TS regions further,
focusing on when they emerge, and what happens when the quasiperiodically modulated quantum wire is
connected with other wires with different modulation phases or an unmodulated one.
Particularly, we would like to
(i) clarify the correspondence between the energy spectrum of the single particle states and emergence of the TS states,
in the presence of either a quasiperiodic modulation with a general wavenumber or the external magnetic field to a general direction,
(ii) understand the effect of phase jumps of the quasiperiodic modulation
and that of a quasiperiodic modulation applied to only a limited part of the one-dimensional system,
and (iii) study the effect of a quasiperiodic modulation of the \textit{hopping} parameter.

In section II we define our model and introduce our Bogoliubov-de Gennes (BdG) equation approach to study the system.
We find that with a lattice-site energy modulation with a generic wavenumber, quasiperiodic or periodic, new topological superconducting regions with end Majorana fermions emerge.
We also study the effect of the direction of the external magnetic field.
Then in section III we observe that those regions are stable against phase jumps in one-dimensional systems,
while when the modulated wire is connected with an unmodulated wire, the locations of the localized modes are determined by which of the wires becomes TS.
We also study the case with quasiperiodic modulation in the intersite hopping parameter,
and find that multiple topological transitions into and out of TS states occur.
Finally, in section IV, we summarize our findings.

\section{Site level modulation by a single (quasi)periodic lattice potential}
We consider a one-dimensional quantum wire parallel to the $\hat{\ve{x}}$ direction,
coupled to a bulk superfluid whose surface is perpendicular to $\hat{\ve{z}}$.
We study a tight-binding one-dimensional model of spin-$1/2$ fermions with
the Rashba-type spin-orbit coupling,
the mean-field coupling to the bulk superconductor,
and the Zeeman energy due to the external magnetic field $\ve{B}$.

The Hamiltonian we have adopted is
\begin{eqnarray}
\mathcal{H}
&=& -\frac{t}{2}\sum_{l=0}^{L-2}\sum_{\sigma=\uparrow,\downarrow}
(\hat c_{\sigma,l}^\dag \hat c_{\sigma,l+1} + \mathrm{h.c.})\nonumber\\
&+& \frac{\alpha}{2} \sum_{l=0}^{L-2} 
\left(
(\hat c_{\downarrow,l}^\dag \hat c_{\uparrow,l+1}
-\hat c_{\uparrow,l}^\dag \hat c_{\downarrow,l+1}) + \mathrm{h.c.}\right)\nonumber\\
&+& \sum_{l=0}^{L-1} \left(\Delta(\hat c_{\uparrow,l} \hat c_{\downarrow,l} + \mathrm{h.c.})
+\frac{2\Gamma}{\hbar}\hat{\ve{B}}\cdot\ve{S}_l\right)\nonumber\\
&+& \sum_{l=0}^{L-1} \sum_{\sigma=\uparrow,\downarrow}(t-\mu+\epsilon_{\sigma,l}) \hat n_{\sigma,l}
,
\label{eqn:Hamiltonian_gen}
\end{eqnarray}
with $\Gamma \equiv \frac{g\mu_B}{2}B$.
We set $t=1$ as the unit of energy and set $\hbar = 1$ in the following.
In the case of $\hat{\ve{B}} = \hat{\ve{z}}$ the third line becomes
\begin{eqnarray}
\sum_{l=0}^{L-1} \left(\Delta(\hat c_{\uparrow,l} \hat c_{\downarrow,l} + \mathrm{h.c.})
+\Gamma(\hat n_{\uparrow,l}-\hat n_{\downarrow,l})\right).\nonumber
\end{eqnarray}
Here, $\hat c_{\sigma,l}$ annihilates a fermion with spin $\sigma(=\uparrow, \downarrow)$ at site
$l (=0, 1, \ldots, L-1)$,
$\hat n_{\sigma,l} \equiv \hat c^\dag_{\sigma,l} c_{\sigma,l}$,
$t$ determines the nearest-neighbor hopping,
$\alpha$ is the Rashba-type SOI,
$\Delta$ is the coupling to the bulk superconductor,
$\Gamma$ is the Zeeman energy,
$\mu$ is the chemical potential, and
$\epsilon_{\sigma,l}$ is the site energy for spin $\sigma$ on site $l$.

In this work we limit our discussion to the case with $\epsilon_{\sigma,l} = \epsilon_l$.
In the following we introduce a quasiperiodic modulation to the site energy,
and study the energy distribution of the single particle state and its
correspondence with the realization of Majorana end modes.

\begin{figure}
\includegraphics[width=8.66cm]{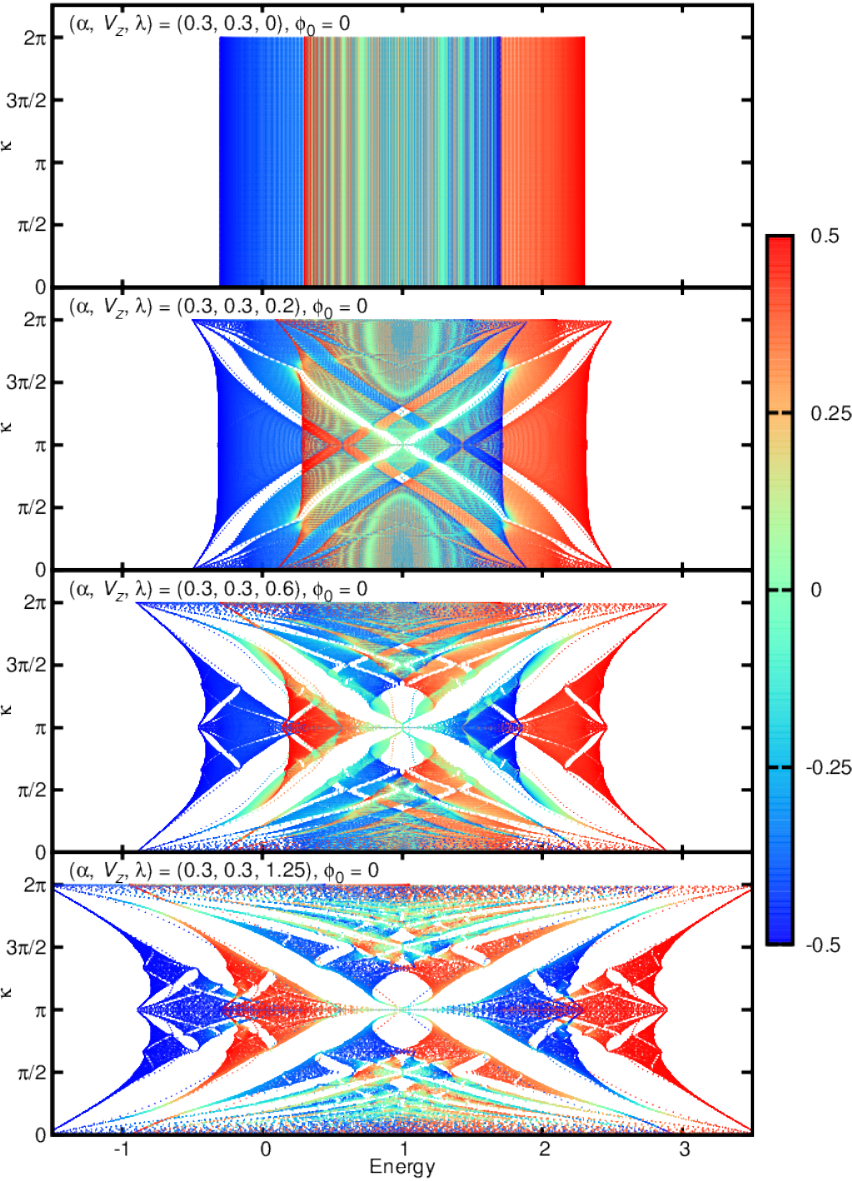}
\caption{(Color online)
The distribution of single particle state energy of eq.~(\ref{eqn:Hamiltonian_gen})
for $L = 200$, $(\Gamma, \alpha)=(0.3, 0.3)$, $V_\mathrm{Q}=0, 0.2, 0.6, 1.25$,
$\hat{\ve{B}} = \hat{\ve{z}}$.
The single particle Hamiltonian, obtained by setting $\Delta = 0$ and $\mu=0$
in eq.~(\ref{eqn:Hamiltonian_gen}), is diagonalized for $\kappa = 2\pi j/400$ ($j=0,1,.\ldots,400$)
and each eigenstate is plotted in a color corresponding to the expectation value of
the $z$ component of the spin for that eigenstate.
The plot for $V_\mathrm{Q} = 1.25$ is similar to that for $V_\mathrm{Q} = 0.8$ (not shown)
except that the former is scaled horizontally,
because of the duality of the lattice between $V_\mathrm{Q} = x$ and $1/x$.
$V_\mathrm{Q} = 1$ is the self-similar point, at which the modulation potential
equals the unmodulated band width for $\Gamma = \alpha = 0$.
}\label{fig:DoubleHofstadter}
\end{figure}

\subsection{Double Hofstadter butterfly}
Let us first consider the case of $\hat{\ve{B}} = \hat{\ve{z}}$.
For an infinitely long system having $\Delta=\epsilon_{l}=0$
we easily obtain the single particle energy as a function of the quasimomentum $k$,
\begin{equation}
E^{\pm}(k) = t(1-\cos(k)) \pm \sqrt{\alpha^2\sin^2(k) + \Gamma^2}.
\label{eqn:RZbands}
\end{equation}
We call them the upper and lower Rashba--Zeeman (RZ) bands. \cite{Tezuka2012}
The mapping of the Hamiltonian to that of a spinless system is possible
if $\mu$ lies in only one of the RZ bands. \cite{Lutchyn2010, Oreg2010, Stoudenmire2011}
In such a case, by introducing the pairing $\Delta$
such that $|\Delta| \lesssim \Gamma$, we obtain the topological superconductor phase.

We consider a site potential which is given by
\begin{equation}
\epsilon_{l} = V_\mathrm{Q} \cos( \kappa (l-l_\textrm{c}) + \phi_0),
\label{eqn:sitePotential}
\end{equation}
in which $V_\mathrm{Q}\geq 0$, $l_\textrm{c} \equiv (L-1)/2$, $\phi_0$ is the phase of the potential
at the center of the system and $\kappa = 2 \pi g$, in which $g$ is a real number such that $0<g<1$.
The potential is periodic for a rational $g$, while it is quasiperiodic for an irrational $g$.
In the following we choose $\phi_0 = 0$, except when we study the effect of $\phi_0$
and when we study the effect of phase jumps in the system.

In a finite length system with $L$ lattice sites, we numerically obtain the set of single particle level energies.
For $\alpha = \Gamma = 0$, each single body wavefunction is extended for $V_\mathrm{Q}<t$ and
localized for $V_\mathrm{Q}>t$ in the $L\rightarrow\infty$ limit. \cite{Kohmoto1983}
The self-similar structure of the two-dimensional spectrum plotted against $\kappa$ for $V_\mathrm{Q} = t$
is called the Hofstadter butterfly. \cite{Hofstadter1976, Kohmoto1983}

When the $2L$ energy levels obtained are plotted against various values of $\kappa$,
for $V_\mathrm{Q} \sim t$, the spectrum shows a self-similar structure resembling
two Hofstadter butterflies shifted in energy and braided together, as shown in Fig.~\ref{fig:DoubleHofstadter}.
We call this structure the double Hofstadter butterfly. \cite{Tezuka2012}

The spin-orbit coupling $\alpha$ mixes the spin-up states and spin-down states differently
at each value of the quasimomentum $k$ of the resulting RZ bands.
Therefore the spin-independent site potential,
which has components with $|k|=\pm \kappa$ and $|k| = \pm(2\pi-\kappa)$,
further mixes the upper and lower RZ bands.
Most of the states in the resulting double Hofstadter butterfly do not have a completely polarized spin.
We may, however, obtain the expectation value of the $z$ component of the spin, $\langle S_z\rangle$,
for each of the $2L$ eigenstate.
The energy levels plotted in Fig.~\ref{fig:DoubleHofstadter} have been color-coded
according to the value of $\langle S_z\rangle$.

We find that, for a fixed value of $\kappa$,
the sets of states from two Hofstadter-butterfly-like structures with
separated values of spin polarizations overlap within some energy ranges.
In some regions in energy there are no single particle states, even inside
the range of $|\epsilon - t|\leq t + \Gamma$, which was occupied by states of
eq.~(\ref{eqn:RZbands}) before the introduction of the site potential modulation
eq.~(\ref{eqn:sitePotential}).
Other regions are occupied by states in only one of the Hofstadter-butterfly-like structure.
We study the consequences of the site potential modulation on the realization of TS states
for the many-body states with a finite chemical potential in the following.

\subsection{Bogoliubov-de Gennes equation: zero modes and Majorana fermions}

The Hamiltonian (\ref{eqn:Hamiltonian_gen}) is bilinear in
operators $\hat{c}$ and $\hat{c}^\dag$.
It can be exactly diagonalized in the Nambu spinor space
$\{ (\ve{u}_\uparrow, \ve{u}_\downarrow, \ve{v}_\uparrow, \ve{v}_\downarrow)^\mathrm{T} \}$,
with the basis obtained as the set of the eigenvectors of the Bogoliubov-de Gennes (BdG) equation.

Alternatively, fermion pairing via a short-range attractive interaction
can also be simulated by introducing a pairing constant $g$ and solving
the Bogoliubov-de Gennes equation self-consistently.
In a lattice system the energy cutoff which renormalizes $g$ can be in principle determined from the lattice constant.
However, here we fix the value of $\Delta$, a homogeneous proximity pairing, by hand.
This is because we do not expect that the BdG approximation,
which simulates the pair formation \textit{within} the 1D wire, 
directly corresponds to our Hamiltonian (\ref{eqn:Hamiltonian_gen}).

We solve \cite{Iskin2012, Liu2012}
\begin{equation}
\left(\begin{array}{cccc}
\hat{H_{\uparrow\uparrow}} & \hat{H_{\uparrow\downarrow}} & 0 & \Delta\\
\hat{H_{\downarrow\uparrow}} & \hat{H_{\downarrow\downarrow}} & -\Delta & 0\\
0 & -\Delta & -\hat{H_{\uparrow\uparrow}} & -\hat{H_{\uparrow\downarrow}}\\
\Delta & 0 & -\hat{H_{\downarrow\uparrow}} & -\hat{H_{\downarrow\downarrow}}\\
\end{array}\right)
\Psi = \epsilon \Psi,
\label{eqn:BdG}
\end{equation}
in which
$\Psi = (\ve{u}_\uparrow, \ve{u}_\downarrow, \ve{v}_\uparrow, \ve{v}_\downarrow)^\mathrm{T}$
is a $4L$-dimensional vector and
$\hat{H_{\sigma\sigma'}}$ are the single-particle components of the Hamiltonian,
\begin{eqnarray}
\hat{H_{\sigma\sigma}} = -\frac{t}{2}\sum_{l=0}^{L-2}\sum_{\sigma=\uparrow,\downarrow}
(\hat c_{\sigma,l}^\dag \hat c_{\sigma,l+1} + \mathrm{h.c.})\nonumber\\
+\sum_{l=0}^{L-1} \sum_{\sigma=\uparrow,\downarrow}(t-\mu+\epsilon_{l}
+(-)^\sigma (\hat{\ve{B}}\cdot\hat{\ve{z}})\Gamma
) \hat n_{\sigma,l},
\end{eqnarray}
\begin{eqnarray}
\hat{H_{\downarrow\uparrow}} &=& \hat{H_{\uparrow\downarrow}}^\dag =
\frac{\alpha}{2} \sum_{l=0}^{L-2} 
(\hat c_{\downarrow,l}^\dag \hat c_{\uparrow,l+1}
-\hat c_{\downarrow,l+1}^\dag \hat c_{\uparrow,l})\\
&-&((\hat{\ve{B}}\cdot\hat{\ve{x}}) + i(\hat{\ve{B}}\cdot\hat{\ve{y}}))\Gamma
\sum_{l=0}^{L-1}\hat c_{\downarrow,l}^\dag \hat c_{\uparrow,l},
\end{eqnarray}
in which $(-)^\sigma = \delta_{\sigma\downarrow}-\delta_{\sigma\uparrow}$.
In the following we call the $4L$-dimensional matrix in the left hand side of eq.~(\ref{eqn:BdG}) $M_\mathrm{BdG}$.

We work in the limit of low temperature $T\rightarrow 0$.
We obtain the particle distribution by
\begin{eqnarray}
n_{\sigma,l} &=& \sum_q\left[f(\epsilon_q) |u_{\sigma,l}|^2 + f(-\epsilon_q) |v_{\sigma,l}|^2\right],
\label{eqn:density}
\end{eqnarray}
in which $f(\epsilon)\equiv (1+\mathrm{e}^{\epsilon/(k_\mathrm{B}T)})
\rightarrow \Theta(-\epsilon)$ ($T\rightarrow 0$) is the Fermi distribution function,
with $\Theta(x)$ being the step function.
The total number of fermions with spin $\sigma$ is obtained as $N_\sigma = \sum_l n_{\sigma,l}$.

The sum in eq. (\ref{eqn:density}) is taken over all $q$.
The matrix $M_\mathrm{BdG}$ is Hermitian, and has pairs of positive and negative eigenvalues with equal absolute values,
because if
$M_\mathrm{BdG}(\ve{u}_\uparrow, \ve{u}_\downarrow, \ve{v}_\uparrow, \ve{v}_\downarrow)^\mathrm{T}
= \epsilon(\ve{u}_\uparrow, \ve{u}_\downarrow, \ve{v}_\uparrow, \ve{v}_\downarrow)^\mathrm{T}$,
we have
$M_\mathrm{BdG}(\ve{v}_\uparrow, \ve{v}_\downarrow, \ve{u}_\uparrow, \ve{u}_\downarrow)^\mathrm{T}
= -\epsilon(\ve{v}_\uparrow, \ve{v}_\downarrow, \ve{u}_\uparrow, \ve{u}_\downarrow)^\mathrm{T}.$
For a positive (negative) eigenvalue, only $|v_{\sigma,l}|^2$ ($|u_{\sigma,l}|^2$) contributes to the particle distribution in the $T\rightarrow0$ limit.

\begin{figure}
\includegraphics[width=8.66cm]{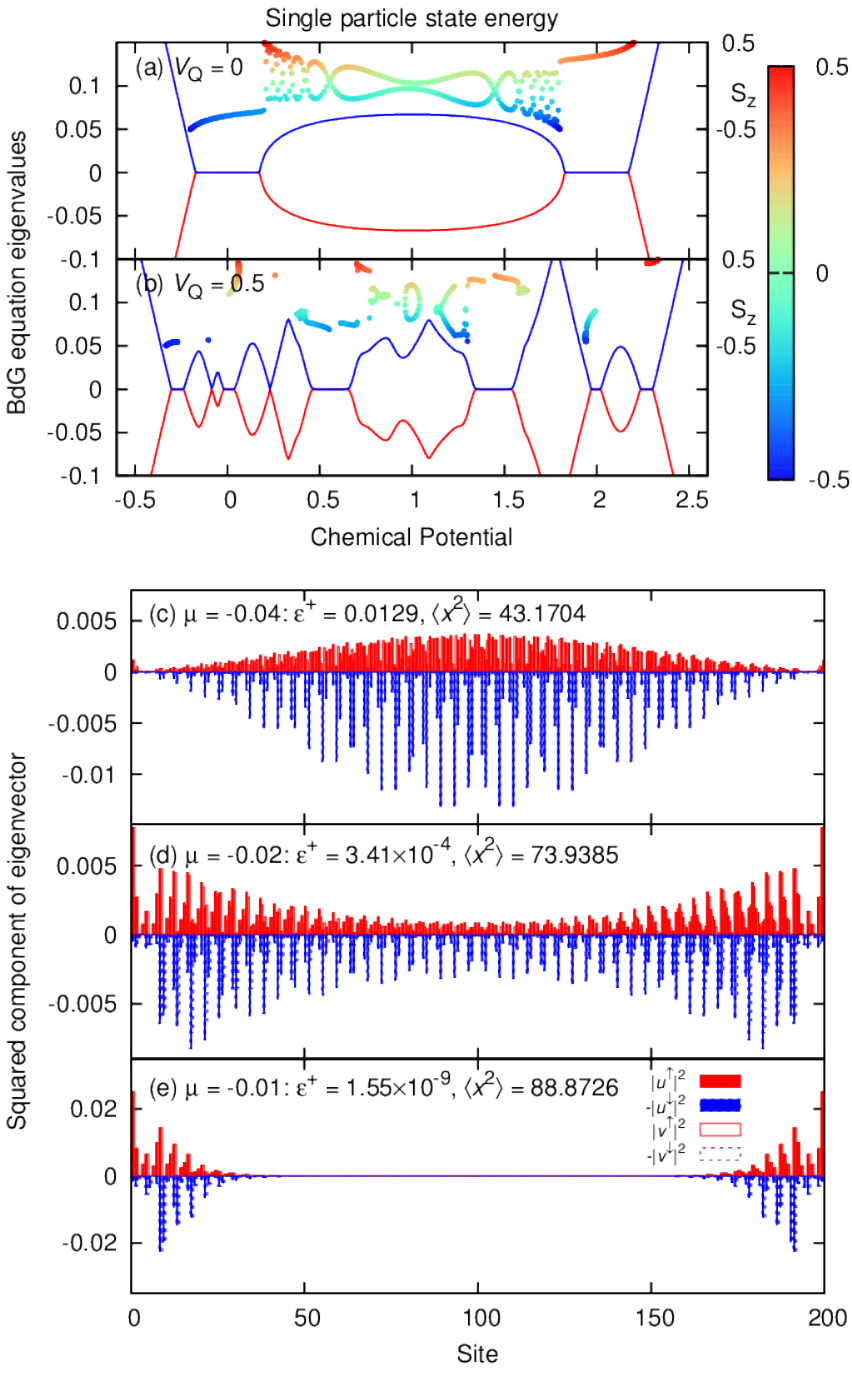}
\caption{(a--b) Largest negative ($-$) and smallest positive ($+$) eigenvalues of the BdG equation
(\ref{eqn:BdG}) plotted against the chemical potential $\mu$ for
$g = \sqrt{5}-2$,
$(\Gamma, \alpha, \Delta)=(0.2, 0.3, 0.1)$ and $V_\mathrm{Q} = $(a) $0$, (b) $0.5$.
(c--e) Amplitudes of the particle- and hole-like parts of the eigenvectors
$\sum_\sigma|u_\sigma|^2$, $\sum_\sigma|v_\sigma|^2$
plotted against the lattice site for 
$(\Gamma, \alpha, \Delta)=(0.2, 0.3, 0.1)$ and $\mu = $(c) $-0.04$, (d) $-0.02$, (e) $-0.01$.}
\label{fig:localization}
\end{figure}

\subsubsection{Distribution of eigenvalues}
For $L$ sites in the system, because of the spin and particle--hole degrees of freedom,
we have $4L$ eigenstates of the BdG equation (\ref{eqn:BdG}).
The introduction of $\Delta$ opens a gap in the eigenvalue spectrum of
eq.~(\ref{eqn:BdG}) in the absence of the spin-orbit coupling $\alpha$ or
the Zeeman field $\Gamma$.
With the spin-orbit coupling and the Zeeman field,
when the chemical potential $\mu$ satisfies 
\[
\mathrm{(lower\ band\ bottom)} < \mu < \mathrm{(upper\ band\ bottom)}
\]
so that the Kitaev model \cite{Kitaev2001} is effectively realized in the spinful case,
the system is a topological superconductor for $0 < \Delta \lesssim \Gamma$ when
$\ve{B}$ is in the $z$--$x$ plane.\cite{Sau2010, Alicea2010, Lutchyn2010, Oreg2010, Stanescu1302.5433}
In this case the BdG equation has two eigenstates with $\epsilon\sim 0$.

Let us consider the $(2L)$-th and $(2L+1)$-th smallest eigenvalues, $\epsilon^-$ and $\epsilon^+$,
which satisfy $-\epsilon^- = \epsilon^+$
because eigenvalues appear in pairs with same absolute value and opposite signs as mentioned above.
We can only have an even number of vanishing eigenvalues, and if we have them $\epsilon^-$ and $\epsilon^+$ should be included in them;
otherwise $\epsilon^- < 0 < \epsilon^+$.

For $\mu\ll\{-t,\min(\epsilon_{l})\}$, the number of fermions in the system is negligible.
As $\mu$ is increased, $N_\sigma$ increases, with $N_\downarrow > N_\uparrow$ for $\Gamma > 0$.
$\epsilon^+$ initially decreases linearly in $\mu$, reflecting the
linear decrease of the required energy to add a single particle in the system.
When it is closer to zero, however, 
the value of $\epsilon_\pm$ approaches more slowly to zero, especially for a smaller system.
When we fit the decrease by a function of the shape $\exp{(-c\vert\mu-\mu_0\vert)}$,
the exponent $c$ is roughly in proportion to the system size $L$.
This also suggests that the modes corresponding to $\epsilon_\pm$ are spatially localized.

\subsubsection{Detection of end Majorana fermions}
From the eigenvector of $M_\mathrm{BdG}$ corresponding to the eigenvalue $\epsilon^+$, 
$(\ve{u}_\uparrow, \ve{u}_\downarrow, \ve{v}_\uparrow, \ve{v}_\downarrow)^\mathrm{T}$,
we define the averaged separation from the system center
\[
\sqrt{\langle x^2 \rangle} \equiv
\sqrt{
\left(\sum_{\sigma,l} (l-l_c)^2 \vert\ve{v}_{\sigma,l}\vert^2\right)
/
\left(\sum_{\sigma,l} \vert\ve{v}_{\sigma,l}\vert^2\right)
}.
\]
If the modes localize to the system ends,
$\sqrt{\langle x^2 \rangle}$ are close to $l_c$, the maximum value it can take.
Note that, when $\epsilon^-$ and $\epsilon^+$ are numerically degenerate
($\epsilon^+ \lesssim 10^{-12}$ in our work),
any linear combination of the two eigenvectors that correspond to these eigenvalues
would be obtained as the eigenvector corresponding to $\epsilon^+$,
so within our BdG calculation we do not directly observe eigenmodes that are localized at only one of the ends of the system.

In the density-matrix renormalization group (DMRG)
simulations of the same model, \cite{Stoudenmire2011, Tezuka2012}
however, a pair of Majorana modes localized at each end of the system have been obtained.
We believe that the DMRG calculation, with limitation on the entanglement entropy between
the two ends of the system, automatically chooses less entangled degenerate ground states,
which are connected to each other by operating either of the two Majorana operators,
for the subspaces with even and odd numbers of fermions.
We find that the localization of $\ve{u}_{\sigma,l}$ and $\ve{v}_{\sigma,l}$,
in terms of $\sqrt{\langle x^2\rangle}$,
precisely corresponds to the localization of the Majorana operators observed by DMRG \cite{Tezuka2012}.

We have plotted the values of $\epsilon^\pm$ as a function of $\mu$ along with the single particle state energy in
Fig.~\ref{fig:localization} (a--b).
In Fig.~\ref{fig:localization} (a) with $V_\mathrm{Q} = 0$, the overlapping RZ bands are
clearly observed, and $\epsilon^\pm$ vanish only when the chemical potential crosses
only one of the RZ bands.

In Fig.~\ref{fig:localization} (b) with $V_\mathrm{Q} = 0.5$ several regions with vanishing
$\epsilon^\pm$ are found, each corresponding to the energy region with 
states from just one of the two Hofstadter-butterfly-like structure.
The components of the eigenvector corresponding to the eigenvalue $\epsilon^+$
are shown in Fig.~\ref{fig:localization} (c--e)
for values of $\mu$ approaching to one of the regions with vanishing $\epsilon^\pm$.
The localization of the mode is clearly observed with an increase of $\sqrt{\langle x^2 \rangle}$.
In all regions with vanishing $\epsilon^\pm$, we observe a clear localization of the corresponding
eigenvectors of $M_\mathrm{BdG}$.

\begin{figure}
\includegraphics[width=8.66cm]{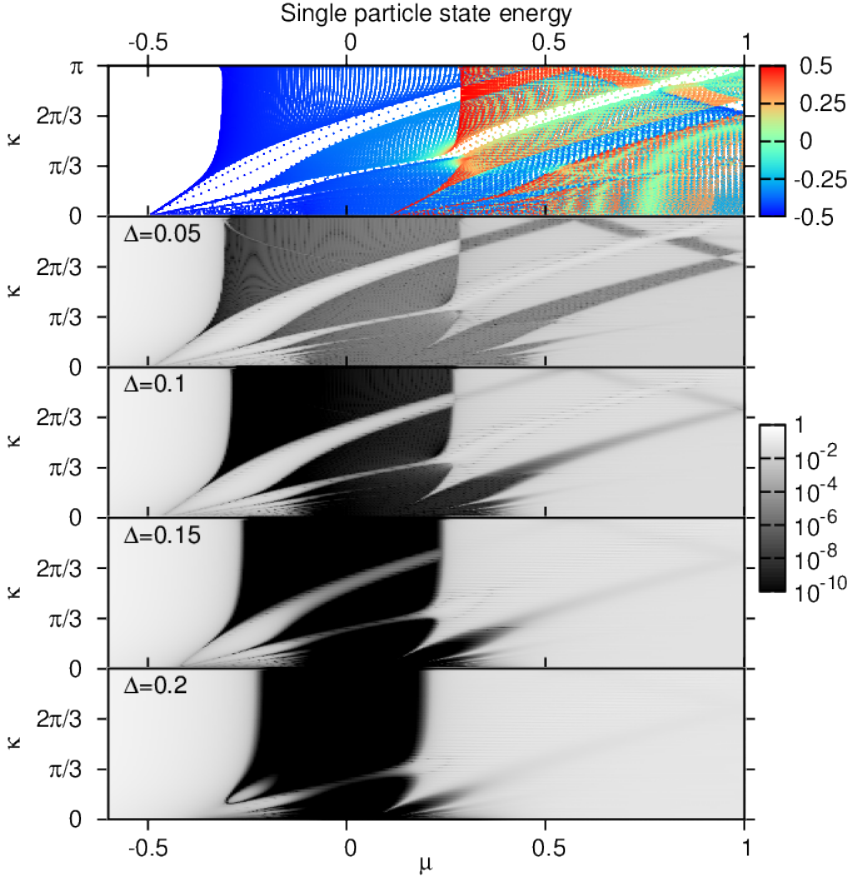}
\caption{(Color online) Top figure: Single particle eigenstates of eq.~(\ref{eqn:Hamiltonian_gen})
color-coded according to the value of $S_z$
for $\hat{\ve{B}}=\hat{\ve{z}}$, $L=400$, $(\Gamma, \alpha, V_\mathrm{Q}) = (0.3, 0.3, 0.2)$ and $-0.6\leq\mu\leq1$, $0 < \kappa < \pi$.
Lower figures: Grayscale plots of $\epsilon^+$ for the same parameters with $\Delta = 0.05, 0.1, 0.15, 0.2$.}
\label{fig:L400lambda0.2}
\end{figure}

\begin{figure}
\includegraphics[width=8.66cm]{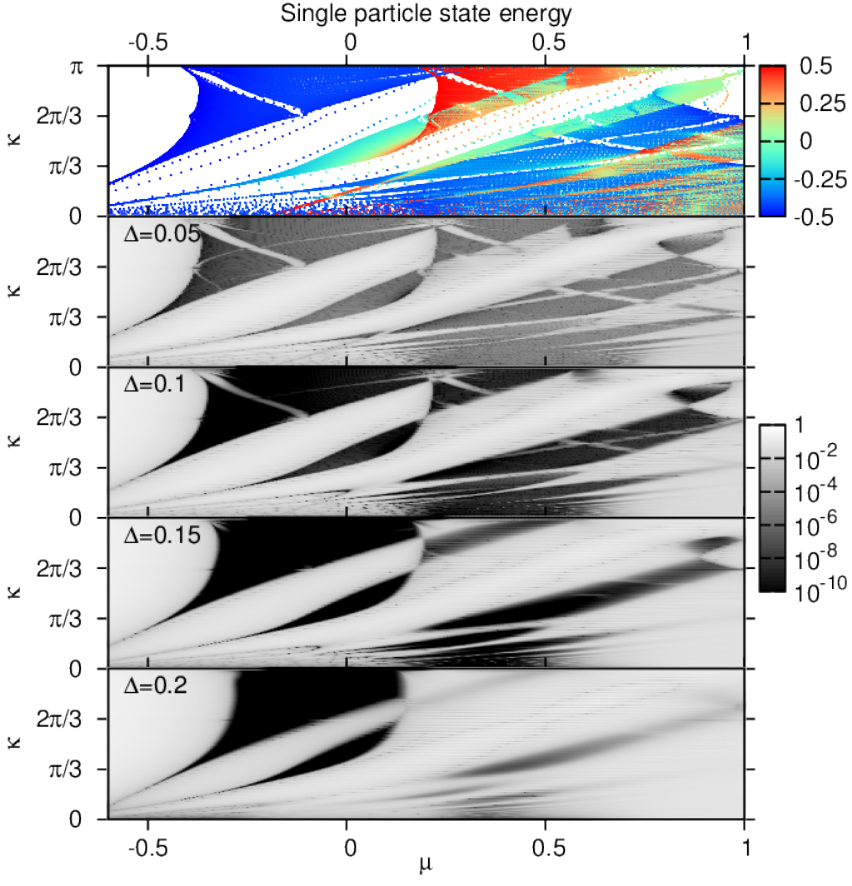}
\caption{(Color online) Top figure: Single particle eigenstates of eq.~(\ref{eqn:Hamiltonian_gen})
color-coded according to the value of $S_z$
for $\hat{\ve{B}}=\hat{\ve{z}}$, $L=400$, $(\Gamma, \alpha, V_\mathrm{Q}) = (0.3, 0.3, 0.5)$ and $-0.6\leq\mu\leq1$, $0 < \kappa < \pi$.
Lower figures: Grayscale plots of $\epsilon^+$ for the same parameters with $\Delta = 0.05, 0.1, 0.15, 0.2$.}
\label{fig:L400lambda0.5}
\end{figure}

\subsection{Dependence on the lattice modulation wavenumber}
In Ref.~\cite{Tezuka2012} we fixed the wave vector $\kappa$
of the quasiperiodic lattice modulation.
However, as we have observed in Fig.~\ref{fig:DoubleHofstadter}, we have effectively
single-band regions of the chemical potential for a wide range of the value of $\kappa$.
Therefore it is interesting to study the dependence of
the appearance of topologically non-trivial superfluid with end Majorana fermions
on the value of $\kappa$.
Especially it is intriguing whether a $\kappa$ such that $2\pi/\kappa$ is an integer
has any difference.

In Figs.~\ref{fig:L400lambda0.2} and \ref{fig:L400lambda0.5}
we plot the value of $\epsilon^+$ in a color code, along with the single particle
state energies obtained from eq.~(\ref{eqn:Hamiltonian_gen}).
For smaller values of $\Delta$, we notice that vanishing eigenvalue corresponds
to regions covered by a single band of single particle eigenstates,
regardless of the value of $\kappa$.
Particularly, even for values of $\kappa$ such as $\pi/3$ and $\pi/2$, $\epsilon^+$ vanishes
when the chemical potential lies in the region where single particle eigenstates are
effectively spinless.

In these regions, because of the effectively single band structure of the non-interacting
Hamiltonian, the Kitaev model is realized, with Majorana fermion modes localizing
at system ends. \cite{Tezuka2012}
The value of $\epsilon^+$ becomes closer to $0$ as $L$ is increased, reflecting that
the end modes are more separated.
For smaller $L$, the separation between the single particle eigenstates is comparable
to $\Delta$ when $\Delta\ll\{\Gamma, t\}$.
In this case, strength of the induced superconductivity depends on the relative
location of a level and the chemical potential.
When the separation is larger (the density of states is lower)
the value of $\epsilon^+$ usually stays larger, but changes rapidly.
$\epsilon^+$ becomes more homogeneous and generally reduced as $\Delta$ is increased
inside each sub-band which does not overlap in energy with another.

We note that, while $N_\downarrow$ is always larger than $N_\uparrow$ for $\Gamma > 0$,
there are sub-bands having fermions almost polarized in the $+\hat{\ve{z}}$ direction
($\langle S_z \rangle \sim 1/2$)
even when $\mu < 1$ with less than half filling factor ($N_\downarrow + N_\uparrow < L$).
Remarkably, even when the chemical potential lies in one of such bands,
a TS state with end Majorana fermions can be formed.

However, as $\Delta$ is increased, smaller features in the single particle state
distribution, which remained in the structure of the value of $\epsilon^+$ for smaller
$\Delta$, gradually disappear and only the widest single-band regions remain visible.
Similar simplification of the phase diagram is also observed in Fig. 2 of
Ref.~\onlinecite{DeGottardi1208.0015} for a quasiperiodic system of spinless fermions.
Finally, for $\Delta \gtrsim \Gamma$, the in-gap state disappears and the
eigenvalue spectrum of $M_\mathrm{BdG}$ has a gap of the order of $\Delta$,
regardless of the value of $V_\mathrm{Q}$,
so that the system is topologically trivial.

In summary, the new topologically non-trivial regions with end Majorana fermions
appear for general site energy modulations of the type of eq.~(\ref{eqn:sitePotential}),
and their existence is not limited to some special irrational values of $g=\kappa/(2\pi)$.
The commensurate case can be considered as a kind of multi-band wire.
The possibility of TS states with end Majorana modes has also been studied
in multi-band systems.\cite{Potter2010,Lutchyn2011a,Rieder2012}
We have observed that while the range of the chemical potential reflects the
single particle eigenstate spectrum strongly, the value of $\Delta$ also plays an important role.
If $\Delta$ is too small $\epsilon^\pm$ vanishes only for regions with higher density of states.
If $\Delta$ is too large, smaller features in the single particle spectrum become smeared.
This occurs before $\Delta$ exceeds $\Gamma$ so that the system becomes topologically trivial
regardless of the values of $\mu$.

\subsection{Dependence on the direction of the external magnetic field}

\begin{figure}
\includegraphics[width=8.66cm]{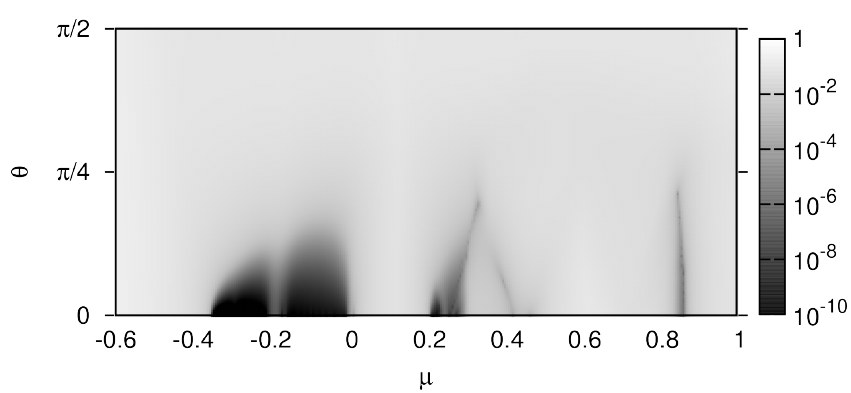}
\caption{(Color online)
Grayscale plot of the value of $\epsilon^+$
for $\hat{\ve{B}}=\hat{\ve{x}}\cos\theta + \hat{\ve{y}}\sin\theta$,
$L=200$, $(\Gamma, \alpha, \Delta, V_\mathrm{Q}) = (0.3, 0.3, 0.1, 0.5)$, $g = (\sqrt5-1)/2$, and $-0.6\leq\mu\leq1$, $0 \leq \theta \leq \pi/2$.
}
\label{fig:rotateTheta}
\end{figure}

The existence of the Majorana end fermions depends on the direction of the applied magnetic field.
\cite{Sau2010, Alicea2010, Lutchyn2010, Oreg2010}
Namely, the effective magnetic field introduced by the Rashba spin-momentum coupling
needs to have perpendicular component to the external magnetic field.
In our model the Rashba spin-momentum coupling is in the $y$ direction, so the
perpendicular directions lie in the $z$--$x$ plane,
Here we ask: can the spin-insensitive quasiperiodic modulation change this situation?

To answer this question we now consider
$\hat{\ve{B}} = \hat{\ve{x}}\cos\theta + \hat{\ve{y}}\sin\theta$,
with $0\leq \theta \leq \pi/2$.
We plot the value of $\epsilon^+$ in a color-coded plot in Fig.~\ref{fig:rotateTheta}.
While the region with vanishing $\epsilon^+$ persists up to $\theta \lesssim \pi/4$,
which corresponds to $\Gamma_x\equiv \Gamma\cos\theta
\gtrsim 0.2$, for larger $\theta$ with $\Gamma_x \lesssim 0.2$
we no longer have a vanishing eigenvalue
of the BdG equation and the system is topologically nontrivial.

Therefore, the introduction of the site level modulation still does not
lift the limitation in the direction of $\ve{B}$ for TS states to be observed.
Because $\Delta = 0.1$ still does not exceed $\Gamma_x$, the result above indicates that 
the $y$ component of the applied magnetic field is rather detrimental
for the realization of a TS in our model.

\subsection{Dependence on the phase of the modulation potential}

\begin{figure}
\includegraphics[width=8.66cm]{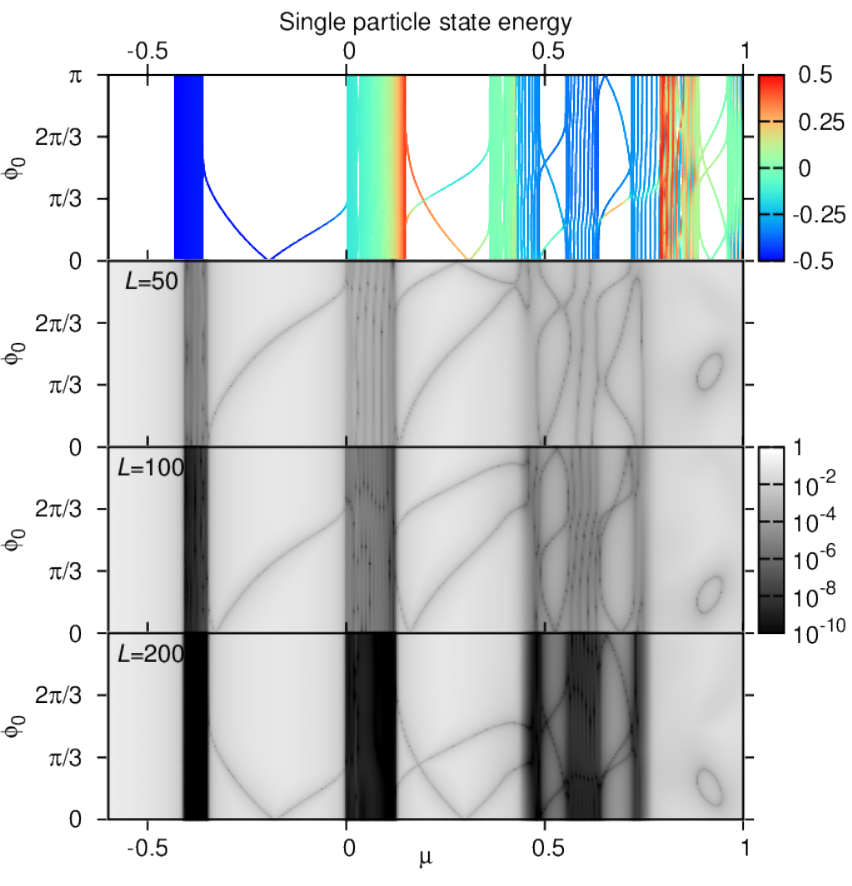}
\caption{(Color online) Top figure: Single particle eigenstates of eq.~(\ref{eqn:Hamiltonian_gen})
color-coded according to the value of $S_z$
for $g=\sqrt5-2$,
$\hat{\ve{B}}=\hat{\ve{z}}$, $L=200$, $(\Gamma, \alpha, V_\mathrm{Q}) = (0.3, 0.3, 0.5)$ and $-0.6\leq\mu\leq1$, $0 \leq \phi_0 \leq \pi$.
Lower figures: Grayscale plots of $\epsilon^+$ for the same set of parameters except that
$\Delta = 0.1$ is introduced and $L=50, 100, 200$.}
\label{fig:f-L200lambda0.5}
\end{figure}

It has been known (see \textit{e.g.} \cite{Mei2012, Ganeshan1301.5639})
for systems with $\Gamma = \alpha = 0$
that when we fix the value of $\kappa$ one or more in-gap states emerge within
the energy gaps due to the potential $\epsilon_{l}$.
Such in-gap states also exist in our model with non-zero $\Gamma$ and $\alpha$,
and can cross each other or another sub-band because the RZ bands are shifted
in energy.
As we change the value of the phase of the site energy modulation, $\phi_0$,
the energies of these in-gap states change rapidly,
while other states in `bulk' sub-bands do not change significantly.

In Fig.~\ref{fig:f-L200lambda0.5} we change the value of $\phi_0$ to study the effect of
such in-gap states.
We have plotted the single particle state energies color-coded by the value of $\langle S_z \rangle$
as well as the value of $\epsilon^+$ for different system sizes.
The plot at the bottom with $L = 200$ looks similar to that of the single particle state energies
at the top, except that the regions with two overlapping sub-bands do not have a vanishing $\epsilon^+$.

We observe that the dependence of the eigenvalues on the choice of
the phase $\phi_0$ becomes weaker as $L$ increases.
Bulk sub-bands are not shifted, and crossing with a single in-gap state
does not usually break the effectively single band situation.
Therefore the topological equivalence between systems with different modulation phases is clear.
In the next section we study what happens if the modulation phase abruptly changes inside the quantum wire,
or if the modulation disappears from a part of the wire.

\section{Effects of different types of lattice modulation}

\subsection{Effect of phase jumps}
In applications to the quantum information field,
namely quantum computation utilizing the pair annihilation or creation of multiple Majorana fermions via gates, \cite{Wu1302.3947}
end Majorana fermions should be stable against minor changes
of the condition of the internals of the one-dimensional system,
which would occur when two or more quantum wires are joined via gates.
Our model of modulated lattices is characterized by the pair of the wavenumber $\kappa$ and
the phase $\phi_0$.
Here, it is of much interest what happens if we have phase jumps of the lattice modulation,
which would correspond to joints between quantum wires, in our system.

Let us consider a system with $N_\mathrm{J}$ phase jumps,
\begin{equation}
\epsilon_{l} = V_\mathrm{Q} \cos( \kappa (l-l_\textrm{c}) + \phi_0 + \lfloor l/W \rfloor \phi_\mathrm{J}),
\label{eqn:sitePotentialJump}
\end{equation}
in which $\lfloor x \rfloor$ denotes the largest integer not exceeding $x$ and
$W = L / (N_\mathrm{J} + 1)$ is the distance between phase jumps of $\phi_\mathrm{J}$.
If the phase jumps do not affect the Majorana modes, the regions of $\mu$ with
vanishing $\epsilon^\pm$ should not change,
and when they vanish, the corresponding eigenmodes should occupy the ends of the system
in spite of the internal phase jumps.

\begin{figure}
\includegraphics[width=8.66cm]{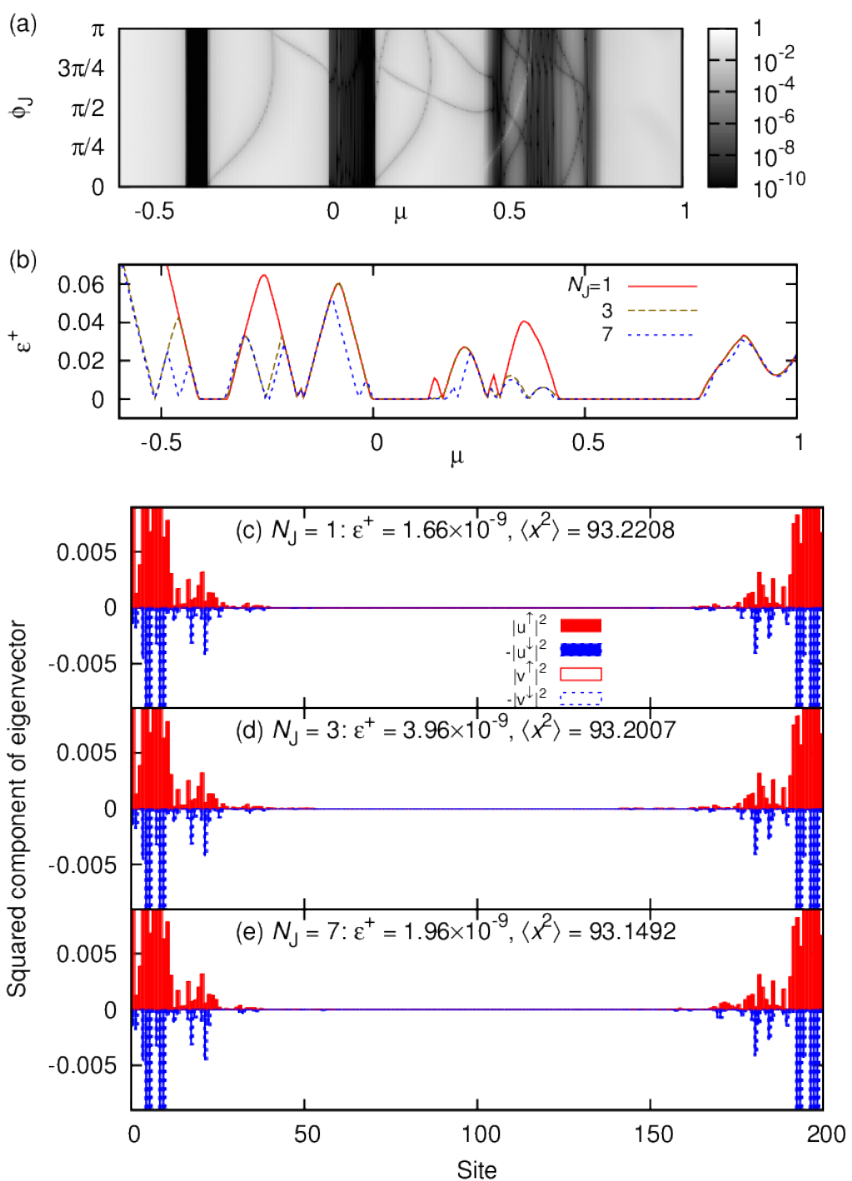}
\caption{(a) Grayscale plots of $\epsilon^+$ for $g=\sqrt5-2$,
$\hat{\ve{B}}=\hat{\ve{z}}$, $L=200$, $(\Gamma, \alpha, V_\mathrm{Q}) = (0.3, 0.3, 0.5)$, $\Delta = 0.1$,
$-0.6\leq\mu\leq1$, $N_\mathrm{J}$ and $0 \leq \phi_\mathrm{J} \leq \pi$.
(b) Value of $\epsilon^+$ plotted against $\mu$ for $N_\mathrm{J} = 1, 3, 7$ and $\phi_\mathrm{J} = \pi$.
Eigenstates for $\mu = 0.04$ are plotted for (c) $N_\mathrm{J} = 1$, (d) $N_\mathrm{J} = 3$, and (e) $N_\mathrm{J} = 7$.
}
\label{fig:phiJ}
\end{figure}

In Fig.~\ref{fig:phiJ} we plot the value of $\epsilon^+$
for (a) a single jump with different sizes of phase jump $\phi_\mathrm{J}$,
and (b) different numbers of phase jumps with $\phi_\mathrm{J} = \pi$.
In Fig.~\ref{fig:phiJ} (a),
$\phi_\mathrm{J} = 0$ corresponds to a system without a phase jump.
Introduction of a single phase jump almost does not change the locations of the regions
with vanishing $\epsilon^\pm$, though we observe a few curves of kinks in $\epsilon^+$
corresponding to a single particle state running between RZ bands as $\phi_\mathrm{J}$ is changed.
Increasing the number of phase jumps does not change the picture significantly,
as long as the localization of the end modes is almost within a single section between phase jumps.
This is observed in Fig.~\ref{fig:phiJ} (b); while the values of $\epsilon^+$ sometimes
differ between systems with different numbers of phase jumps of $\pi$,
the regions with $\epsilon^+\ll 1$ is not shifted or removed.
Also we find in Fig.~\ref{fig:phiJ} (c--e)
that the eigenvalues of $M_\mathrm{BdG}$ other than $\epsilon^\pm$ do not vanish in this case.
There are only one pair of Majorana fermions appearing at the both ends of the system,
rather than more than one pairs of them appearing also at phase jumps.

The results above reflect the fact that a phase jump does not change the topological character of the system.
If both sides of the introduced phase jump are in the state characterized by
the same topological quantum number, boundary states do not form at the phase jump.

\subsection{Site level modulation limited to a part of the system}

\begin{figure}
\includegraphics[width=8.66cm]{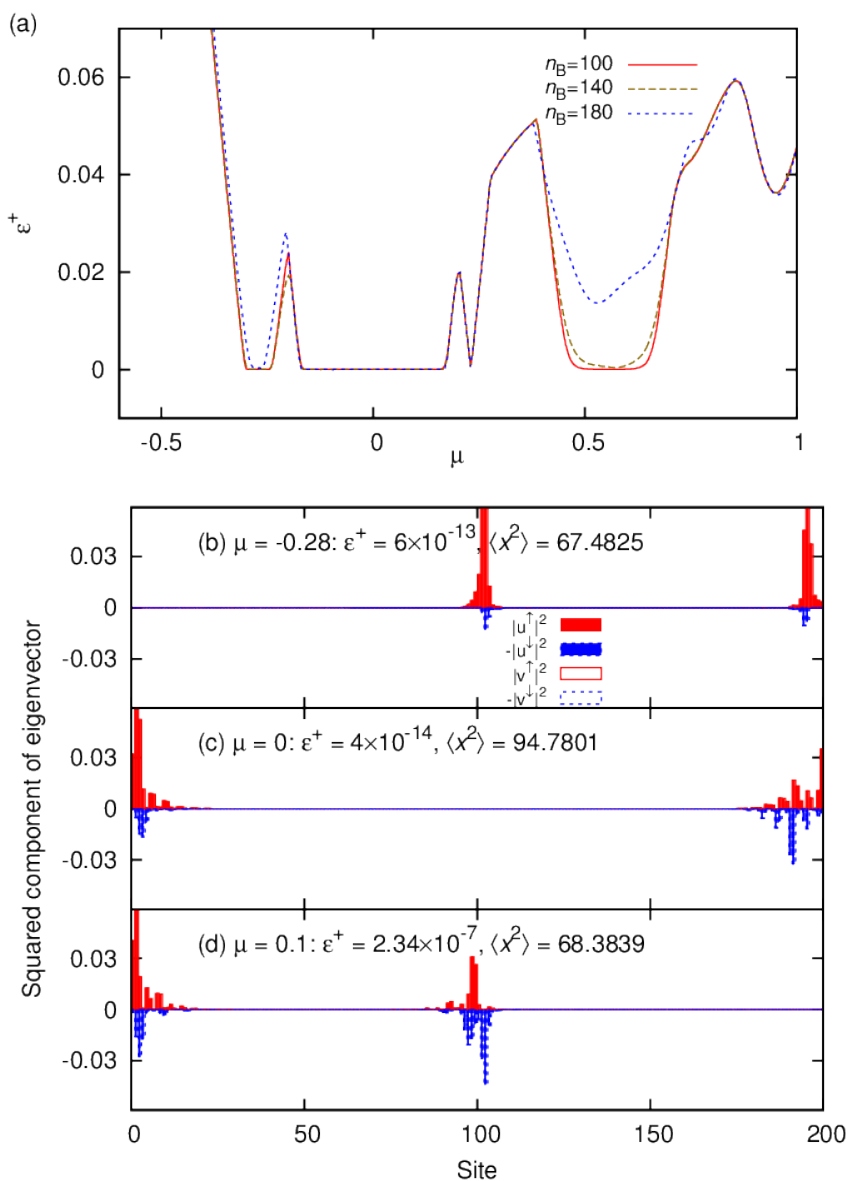}
\caption{
(a) Value of $\epsilon^+$ plotted against $\mu$ for $n_\mathrm{B} = 100, 140, 180$
and $g=\sqrt5-2$,
$\hat{\ve{B}}=\hat{\ve{z}}$, $L=200$, $(\Gamma, \alpha, V_\mathrm{Q}) = (0.2, 0.3, 0.5)$, $\Delta = 0.1$.
Eigenstates for $n_\mathrm{B} = 100$ are plotted for
(b) $\mu = -0.28$ (only the right side of the system is single band),
(c) $\mu = 0$ (both sides are single band),
and (d) $\mu = 0.1$ (only the left side is single band).
}
\label{fig:partialModulation}
\end{figure}

The view above is further confirmed when we remove the lattice modulation from some part of the system,
by having
\begin{equation}
\epsilon_l = \left\{
\begin{array}{cc}
0, & (0 \leq l < n_\mathrm{B})\\
V_\mathrm{Q} \cos(\kappa(l - l_c)). & (n_\mathrm{B} \leq l < L)
\end{array}
\right.
\end{equation}
In Fig.~\ref{fig:partialModulation} (a) we plot the value of $\epsilon^+$ for different values of $n_\mathrm{B}$
for the case in which the part of the system to the left of site $n_\mathrm{B}$ has the same parameter as in
Fig.~\ref{fig:localization} (a), whereas the right part has the same parameter as in Fig.~\ref{fig:localization} (b).
For $n_\mathrm{B} = L/2 = 100$ we observe that $\epsilon^+$ vanishes when it vanishes either
in Fig.~\ref{fig:localization} (a) with $V_\mathrm{Q} = 0$ or (b) with $V_\mathrm{Q} = 0.5$.
In Fig.~\ref{fig:partialModulation} (b--d) we have plotted the spatial distribution of
eigenvectors of (\ref{eqn:BdG}) corresponding to $\epsilon^+$.
In Fig.~\ref{fig:partialModulation} (b) with $\mu = -0.28$, only the right side of the system with
$V_\mathrm{Q} = 0.5$ is an effectively single band system at the chemical potential and
becomes a topological superconductor. The eigenvector has most of its amplitude localized
equally to the two ends of this region, and the value of $\sqrt{\langle x^2 \rangle}$ is close to $L/\sqrt{2}$,
as expected in such a case.
Also, in Fig.~\ref{fig:partialModulation} (d) with $\mu = 0.1$, only the left side of the system
without the lattice modulation becomes a topological superconductor, and the eigenvector
is this time localized to the two ends of the left part, with the value of $\sqrt{\langle x^2 \rangle}$
similar to that in (b).
On the other hand, in Fig.~\ref{fig:partialModulation} (c) both sides are topologically nontrivial,
and the eigenvector has its amplitudes localized at the two ends of the whole system.

As we enlarge the part without lattice modulation by increasing the value of $n_\mathrm{B}$
in Fig.~\ref{fig:partialModulation} (a),
the plot of $\epsilon^+$ approaches that in Fig.~\ref{fig:localization} (a).
This is because the part with the lattice modulation cannot form a well-defined topological superconductor
if it is too short.

In summary,
when two TS regions are joined, the resulting system become a TS with end Majorana states at the ends.
On the other hand, if a TS is joined with a topologically trivial chain,
the resulting system becomes a TS whose Majorana states appear close to where they had been in the original TS.
This does not depend on which of the chains has spatial modulation.
If the chemical potential can be controlled in the system one can control which region has boundary Majorana modes
in the setting above.

\subsection{Case of a hopping modulation}

\begin{figure}
\includegraphics[width=8.66cm]{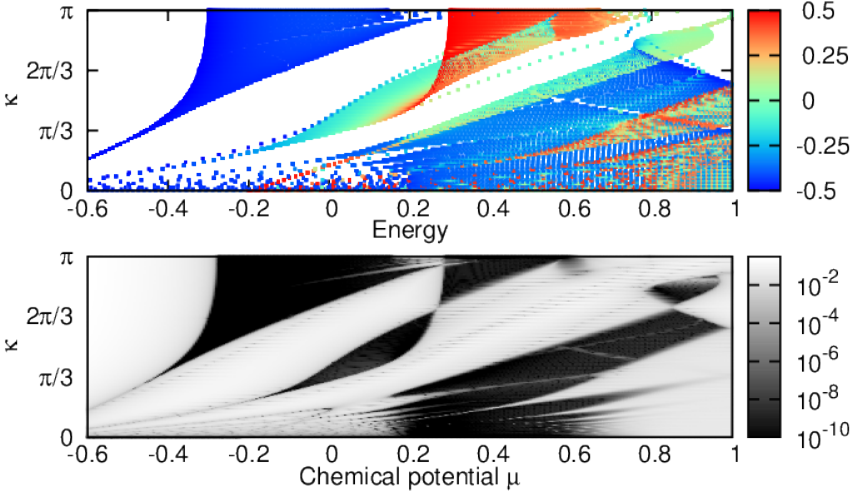}
\caption{(Color online)
Effect of quasiperiodic modulation of the hopping parameter ---
Top: Single particle state energy for $\hat{\ve{B}}=\hat{\ve{z}}$, $L=200$,
$(\Gamma, \alpha, V_\mathrm{J}) = (0.3, 0.3, 0.5)$ and
$-0.6\leq\mu\leq1$, $0 \leq \kappa \leq \pi$.
Bottom: $\epsilon^+$ calculated for the same parameters and $\Delta = 0.1$.
}
\label{fig:hopMod05}
\end{figure}

The correspondence between the `diagonal' modulation in the lattice site energy and
the `off-diagonal' one in the hopping amplitude in quasiperiodic systems
has attracted a renewed attention. \cite{Kraus2012, Thiem1212.6337}
Here we study the effect of such a modulation in the hopping amplitude in our system,
in which case the Hamiltonian is given by
\begin{eqnarray}
\mathcal{H}
&=& -\sum_{l=0}^{L-2}\frac{t_l}{2}\sum_{\sigma=\uparrow,\downarrow}
(\hat c_{\sigma,l}^\dag \hat c_{\sigma,l+1} + \mathrm{h.c.})\nonumber\\
&+& \frac{\alpha}{2} \sum_{l=0}^{L-2} 
\left(
(\hat c_{\downarrow,l}^\dag \hat c_{\uparrow,l+1}
-\hat c_{\uparrow,l}^\dag \hat c_{\downarrow,l+1}) + \mathrm{h.c.}\right)\nonumber\\
&+& \sum_{l=0}^{L-1} \left(\Delta(\hat c_{\uparrow,l} \hat c_{\downarrow,l} + \mathrm{h.c.})
+\frac{2\Gamma}{\hbar}\hat{\ve{B}}\cdot\ve{S}_l\right)\nonumber\\
&+& \sum_{l=0}^{L-1} \sum_{\sigma=\uparrow,\downarrow}(t-\mu+\epsilon_{l}) \hat n_{\sigma,l}
,
\label{eqn:Hamiltonian_offDiagMod}
\end{eqnarray}
with
\begin{equation}
t_l = t\left[1 + V_\mathrm{J} \cos\left(\kappa (l + 1/2 - l_c) + \phi_0\right)\right]
\end{equation}
and $\epsilon_{l} = 0$.

In Fig.~\ref{fig:hopMod05} we plot the single particle state energy
as well as the value of $\epsilon^+$ obtained by solving the BdG equation
similar to (\ref{eqn:BdG}) but the diagonal (spin-preserving) blocks of the kinetic term
substituted by the one with the modulated hopping parameter.
We observe that, while the details of the single particle spectrum is changed,
the correspondence between the effectively single-band region in the spectrum and
the zeros of the $\epsilon^+$ is also observed here.

The symmetry of the Hamiltonian is not changed by going from the site level modulation
to the hopping parameter modulation.
Therefore similar response to phase jumps or partial modulations inside the system
is expected for the latter case as in the former case, which has been studied in the above.

\section{Conclusion}

In summary, we have studied the effect of different types of spatial modulations
on the realization of a topological superconductor in a 1D conductor with
proximity-induced superfluidity and spin-orbit coupling,
extending our previous work, Ref.~\onlinecite{Tezuka2012}.

The combination of a quasiperiodic site energy modulation
with the external Zeeman field and the spin-orbit coupling
results in the single particle state energy distribution having a fractal pattern,
\textit{double Hofstadter butterfly}.
Within the mean-field, Bogoliubov-de Gennes approximation, our model Hamiltonian
can be diagonalized. We have demonstrated that the smallest positive eigenvalue
of the Hamiltonian is strongly governed by the single particle energy spectrum
for relatively weak induced superfluidity. Localized end modes, which are
Majorana fermions, exist when two eigenvalues are degenerate at zero energy.
As we change the chemical potential or the modulation wavenumber, we observe
multiple reentrant transitions into and out of topologically nontrivial states.
However, for stronger superfluidity, small patterns of the double Hofstadter butterfly are smeared
from the eigenvalue plot showing the topologically nontrivial parameter ranges.
The resulting topological superconductor is sensitive to the direction of the magnetic field,
while the phase of the modulation does not affect the system.

We have also studied the effects of the phase jump of the quasiperiodic potential
and what happens when the potential is applied to only a part of the quantum wire.
The results reflect that the system is characterized by a $Z_2$ quantum number,
that is, all the topologically nontrivial states are indistinguishable, and
if two regions with such states are joined, the Majorana end modes appear only
at the ends of the resulting system.
If the chemical potential can be changed, the locations of the end modes can be manipulated.
A quasiperiodic hopping modulation also exhibits a similar phase diagram with reentrant topological transitions.

Recently a scheme for topological superconductivity without proximity effect
has been proposed \cite{Farrell2012}.
Our study of the correlation between the band structure and realization of
topological superconductivity could also be relevant in such cases.

\begin{acknowledgments}
M.~T. acknowledges the hospitality of Cavendish Laboratory,
University of Cambridge, where this work was completed.
This work was partially supported by the Grant-in-Aid for the Global COE
Program ``The Next Generation of Physics, Spun from
Universality and Emergence'' from MEXT of Japan.
N.~K. is supported by KAKENHI (Grants No. 22103005
and No. 25400366) and JSPS through its FIRST Program.
Part of the computation in this work has been performed using the facilities of the Supercomputer Center,
Institute for Solid State Physics, University of Tokyo
and Yukawa Institute for Theoretical Physics, Kyoto University.
\end{acknowledgments}

\end{document}